\documentclass[10pt]{article}
\usepackage{a4wide}
\usepackage{color}
\usepackage{cite}
\usepackage{graphicx}

\usepackage{amssymb}
\usepackage{amsmath}
\usepackage{epstopdf}
\usepackage{soul}

\newcommand{\dd}{{\rm{d}}}
\newcommand{\DD}{{\rm{D}}} 

\newcommand{\im}{\mathrm{i}}
\newcommand{\e}{\mathrm{e}}

\newcommand{\bolde}{\boldsymbol{e}}
\newcommand{\boldk}{\boldsymbol{k}}
\newcommand{\boldl}{\boldsymbol{l}}
\newcommand{\boldm}{\boldsymbol{m}}
\newcommand{\boldu}{\boldsymbol{u}}

\DeclareMathOperator{\sech}{sech}

\begin{document}

\linespread{1.0}

\title{
Interpreting gravitational fields of \\
Topologically Massive Gravity using geodesic deviation
}

\author{
Mat\'u\v{s} Papaj\v{c}\'{\i}k
and
Ji\v{r}\'{\i} Podolsk\'y\thanks{{\tt
matus.papajcik@matfyz.cuni.cz,
podolsky@mbox.troja.mff.cuni.cz,
} }
\\ \ \\ \ \\
Charles University, Faculty of Mathematics and Physics, \\
Institute of Theoretical Physics, \\
V~Hole\v{s}ovi\v{c}k\'ach 2, 18000 Prague 8, Czechia.
}

\maketitle

\begin{abstract}
We study relative motion of nearby test particles in Topologically Massive Gravity (TMG) in three spacetime dimensions, using the equation of geodesic deviation. We show that, in a suitable reference frame, the influence of any gravitational field can be decomposed into transverse, longitudinal, and Newtonian components, which are directly related to the Cotton scalars of the Newman--Penrose-type. In particular, we prove that Cotton type N spacetimes exhibit a purely transverse gravitational effect on test particles, and can thus be reasonably interpreted as specific gravitational waves with a single polarization mode in TMG. The influence of the cosmological constant manifests itself as an isotropic effect. We also discuss the physical interpretation of spacetimes of specific algebraic types, as well as the influence of various matter fields, namely pure radiation, perfect fluid, and electromagnetic field. As an example, we provide an explicit analysis of TN-waves and pp-waves in three-dimensional TMG.
\end{abstract}

\vfil\noindent
PACS class:  04.20.Jb, 04.50.--h, 04.40.Nr, 04.30.Nk 


\bigskip\noindent
Keywords: 3D gravity, Topologically Massive Gravity, geodesic deviation, Cotton scalars, algebraic types, gravitational waves, pp-waves

\vfil
\eject

\tableofcontents

\newpage

\linespread{1.5}

\section{Introduction}

For several decades now, various theories of gravity with massive gravitons (a proposed quantum particle of gravitational interaction) have received a great attention. Such theories are conceptually important, particularly in addressing the cosmological constant problem, or attempting to understand the origin of the late-time acceleration of the universe.

Since the General Relativity (GR) is the unique theory of a massless spin-2 particle \cite{BD:1975}, any modifications to GR must either break the Lorentz invariance, or introduce a massive graviton. According to the most up-to-date experiments \cite{Ligo:2018, BMFGLD:2019}, the mass of the graviton is constrained to be less than ${10^{-23}\, \text{eV/c}^2}$. Unfortunately, Lorentz-invariant massive gravity theories in four dimensions (4$D$) typically suffer from the Boulware--Deser ghosts (for a review see \cite{deRham:2014,Hinterbichler:2021}). This issue was resolved in the seminal work of de~Rham, Gabadadze and Tolley \cite{dRGT:2011}, who formulated a ghost-free theory of massive gravity in 4$D$.

Interestingly, long before the development of this ghost-free theory in 4$D$, lower-dimensional 2+1 gravity models were explored to investigate this issue. In particular, a consistent massive gravity theory was formulated in 1982, when Deser, Jackiw and Templeton modified the standard Einstein--Hilbert action in three spacetime dimensions (3$D$) by adding the Chern--Simons term \cite{DJT:1982a, DJT:1982b}. This 3$D$ theory of gravity, known as the Topologically Massive Gravity (TMG), introduces a dynamical degree of freedom that is otherwise absent in standard 3$D$ GR \cite{DJH:1984, DJ:1984}. Specifically, it describes a massive spin-2 field with a single helicity mode, free of ghosts. However, it also has several drawbacks: To preserve unitarity the scalar curvature must appear with a negative sign in the action, and the Chern--Simons term leads to parity violation. The latter problem was addressed in a more recent model of 3$D$ gravity called the New Massive Gravity \cite{BHT:2009}.

Currently, TMG plays an important role in ongoing research on quantum gravity. Its canonical quantization was performed already in \cite{BLK:1993}. Similar lower-dimensional models were popularized by Witten in 1988 in his ground-breaking work \cite{Witten:1988}. These theories serve as test beds for different approaches to the quantization of gravity. A comprehensive review of the subject is available in the monograph \cite{Carlip:2003}. Additionally, study of specific aspects of the AdS/CFT correspondence and TMG remains an active area of research \cite{GJ:2008,STR:2009,GGV:2010,ACEGJ:2011,AQNS:2016,CHLZ:2016,LR:2020,Kaplan:2023}. Moreover, field-theoretical investigations have drawn interest to the classical relativistic properties of the theory, including asymptotic symmetries of TMG \cite{CMMRRV:2018,AJTYZ:2021}, and most notably the study of its exact solutions. This has led to a systematic analysis of vacuum solutions in \cite{CPS:2010a}, and an investigation of all possible Kundt geometries \cite{CPS:2010b}. An extensive overview of the known exact spacetimes of TMG can be found in Chapters 16--19 of the monograph \cite{Garcia-Diaz:2017}, and in the references therein.

While a significant effort has been devoted to finding exact solutions in TMG, a general approach to their deeper physical interpretation remains lacking. Our work here presents a systematic method for investigation of exact 2+1 spacetimes of TMG, based on the relative motion of test particles. For freely falling, uncharged, and spin-less nearby particles, the motion is described by the well-known equation of geodesic deviation, first derived a century ago by Levi-Civita and Synge \cite{Levi-Civita:1926, Synge:1926, Synge:1934} (for a historical account see \cite{Trautman:2009}). To this day, the equation of geodesic deviation remains a fundamental tool in understanding various gravitational phenomena, including the tidal effects experienced by observers falling into black holes, and the influence of gravitational waves on test bodies, see e.g. \cite{Mashhoon:1975,Mashhoon:1977,Tammelo:1977,AL:1983,CD:1986,BJ:1989,BP:1999,BHK:2000,Biesiada:2003,CM:2006,BGJ:2011,KH:2011,DMA:2015,CCM:2024}, among many other applications. Numerous further works have extended this equation beyond the linear approximation in the separation vector, also accounting for arbitrary relative velocities of test particles and higher-order corrections \cite{Hodgkinson:1972,Bazanski:1977,Manoff:1979,AP:1979,LN:1979,Schutz:1985,Ciufolini:1986,Vanzo:1992,KHC:2001,Manoff:2001,Holten:2002,CM:2002,HFMS:2005,Perlick:2008,Vines:2015,WB:2022}.

In 1965 Szekeres analysed the behaviour of test particles in a generic 4$D$ spacetime \cite{Szekeres:1965}, using the formalism of self-dual bivectors \cite{JEK:1960,Sachs:1961}. He demonstrated that their relative motion is affected by transverse, longitudinal and Newtonian components of the gravitational field. He also related the presence of these components to distinct Petrov algebraic types. The approach became even more refined when it was re-expressed in the Newman--Penrose formalism \cite{NP:1962,PR:1984} utilizing the complex Weyl scalars $\Psi _A$. The equation of geodesic deviation allows the interpretation of these scalars as important physical quantities, such as $\Psi _4$, which is related to the amplitude of a gravitational wave \cite{BP:1999}. More details of this approach in 4$D$ can be found in monographs \cite{SKMHH:2003,GP:2009}. With the development of the Newman--Penrose-type formalism in higher dimensions \cite{CMPP:2004,KP:2006,Coley:2008}, an analogous analysis of spacetimes of any dimension ${D\geq 4}$, based on the equation of geodesic deviation, was performed in \cite{PS:2012}.

Natural extensions of the Newman--Penrose formalism into lower dimensions face an immediate obstacle, namely that in 3$D$ the Weyl tensor identically vanishes. However, Barrow, Lurd and Lancaster suggested to employ the Cotton tensor $C_{abc}$ as the key object in determining the algebraic structure of 2+1 geometries \cite{BBL:1986}. This method, later refined in \cite{GHHM:2004}, relies on the Jordan decomposition of the Cotton--York tensor $Y_{ab}$. In TMG the Cotton--York tensor is simply related to the Ricci tensor $R_{ab}$, so that the algebraic classification can be shown to be equivalent to that of the traceless Ricci tensor \cite{CPS:2010a,CPS:2010b}. More recently, we developed an analogue to the Newman--Penrose formalism in 3$D$ \cite{PP:2023, PP:2024}, based on five real scalars $\Psi _A$ constructed directly from the Cotton tensor. These scalars encode all information contained in the Cotton tensor, and the algebraic classification of 2+1 spacetimes is determined by the gradual vanishing of $\Psi _A$ in a suitable frame. We have also shown that this method is equivalent to the previous approaches to algebraic classification.

Our paper is organized as follows. Section~2 provides an overview of TMG, including some important properties and the field equations. We derive an equivalent formulation of the field equations useful for further investigations. Section~3 focuses on the fundamental equation of geodesic deviation. We present its coordinate form, and discuss the importance of projecting it onto an observer's reference frame. Using the field equations, we express the Riemann tensor in terms of the Cotton--York and energy-momentum tensors. We then use the Cotton scalars $\Psi _A$ to rewrite the frame components of the Riemann tensor, and formulate a frame component version of the geodesic deviation. Section~4 examines the influence of specific gravitational components, represented by the cosmological constant~$\Lambda $ and the scalars~$\Psi _A$, on the motion of test particles. Section~5 constructs a unique interpretation frame adapted to the given structure of spacetime. This allows us to invariantly interpret spacetimes of any algebraic type. We show that only type N spacetimes exhibit purely transverse effects. Section~6 analyses the effect of pure radiation, perfect fluid, and electromagnetic field. Finally, Section~7 investigates the geodesic deviation for a general class of TN-wave and pp-wave spacetimes in TMG. We derive an explicit solution for the separation vector in a particular pp-wave spacetime, and discuss its physical interpretation.

\section{Topologically Massive Gravity}

It is well known that General Relativity in 3$D$ suffers from the absence of dynamical degrees of freedom, which makes the theory (locally) trivial in empty space. In 1982, Deser, Jackiw and Templeton proposed a modification of the standard Einstein--Hilbert action \cite{DJT:1982a, DJT:1982b} to contain an additional Chern--Simons term,
\begin{equation}
S_{\text{CS}} = \int \dd ^3x \, \sqrt{-g}\, \epsilon ^{abc}\, \Gamma {^m}_{an}\Big( \partial _b\, \Gamma {^n}_{mc}+\frac{2}{3}\,\Gamma {^n}_{bk}\, \Gamma {^k}_{cm} \Big) \, .
\end{equation}
This term, constructed entirely using the Levi-Civita tensor\footnote{We use the definition of the Levi-Civita tensor ${\epsilon ^{abc} \equiv \varepsilon ^{abc} / \sqrt{-g}}$, where $\varepsilon ^{abc}$ is the Levi-Civita symbol.} and the connection coefficients $\Gamma {^a}_{bc}$, introduces a single massive dynamical degree of freedom. It is closely related to topology, as it can be shown \cite{DJT:1982b} to represent a metric-independent scalar proportional to the Chern--Simons secondary characteristic class \cite{Chern:1979}. The action of thus modified theory of gravity in 3$D$ (known as TMG) including the cosmological constant $\Lambda $, has the form\footnote{The cosmological constant was included into the action by Deser in \cite{Deser:1984}.} \cite{DJT:1982a, DJT:1982b}
\begin{equation} \label{Action_TMG}
S = \frac{1}{16\pi} \int \dd ^3x \, \sqrt{-g}\, \Big[\sigma R-2\Lambda +\frac{1}{2\mu}\, \epsilon ^{abc}\, \Gamma {^m}_{an}\Big( \partial _b\, \Gamma {^n}_{mc}+\frac{2}{3}\,\Gamma {^n}_{bk}\, \Gamma {^k}_{cm} \Big) \Big] \, ,
\end{equation}
where $\sigma $ is a discrete constant coupled to the Ricci scalar $R$, which can take the values ${\sigma = \pm 1}$. The choice ${\sigma = 1}$ corresponds to the standard gravitational action, however, it leads to massive gravitons with a negative kinetic energy \cite{LSS:2008, GJ:2008}. Therefore, to retain the unitarity of the theory the opposite sign ${\sigma = -1}$ is often considered. The coupling constant of the Chern--Simons term $\mu $ is a mass parameter of the theory. In the \emph{linearized} regime the action \eqref{Action_TMG} describes a massive spin-2 field (massive graviton) \cite{DJT:1982a, DJT:1982b}, whose mass $m_g$ was identified in \cite{GST:2010} as
\begin{equation}
m_g ^2 = \sigma ^2 \mu ^2 + \Lambda \, .
\end{equation}
In the Minkowski limit (${\Lambda = 0}$) the constant $\mu$ is equal to the mass of the graviton, and for this reason it is regarded as a mass parameter of the theory. Additionally, the Chern--Simons term is responsible for parity breaking, and under the transformation ${\mu \rightarrow -\mu}$ the single helicity mode of the graviton is reversed \cite{GST:2010}.

Variation of the action \eqref{Action_TMG}, assuming also standard coupling to matter sources, leads to the following field equations of TMG \cite{DJT:1982a, DJT:1982b, Garcia-Diaz:2017},
\begin{equation} \label{TMG_field_eqs}
\sigma \, \Big( R_{ab}-\frac{1}{2}R\, g_{ab} \Big) + \Lambda \, g_{ab} + \frac{1}{\mu}\, Y_{ab} = 8\pi \, T_{ab} \, ,
\end{equation}
with $T_{ab}$ denoting the energy-momentum tensor, and $Y_{ab}$ denoting the Cotton--York tensor. In 3$D$ the Cotton--York tensor is a Hodge map of the third-rank Cotton tensor $C_{mnb}$. This tensor plays the role of the conformally invariant tensor in 3$D$, since the Weyl tensor identically vanishes in this lower dimension. It is defined as
\begin{equation} \label{Cotton_tensor}
C_{mnb} \equiv 2 \Big( \nabla _{[m}R_{n]b} - \frac{1}{4}\nabla _{[m}R\, g_{n]b} \Big ) \, .
\end{equation}
It is antisymmetric in the first two indices, ${C_{(mn)b}=0}$, traceless, ${C{_{mn}}^m=0}$, and it satisfies the relation ${C_{[abc]}=0}$. These constraints restrict the number of its independent components to \emph{five} in 3$D$. Therefore, it can be mapped onto a symmetric (${Y_{[ab]}=0}$) and traceless (${Y{^a}_{a}=0}$) second-rank Cotton--York tensor
\begin{equation} \label{Cotton-York_tensor}
Y_{ab} \equiv \tfrac{1}{2} \, g_{ak} \, \epsilon ^{kmn} \, C_{mnb} \, .
\end{equation}
Because this mapping is one-to-one, the Cotton--York tensor encodes the same information as the Cotton tensor. Specifically, it is a conformally invariant tensor, and it vanishes for conformally flat 3$D$ spacetimes. The presence of derivatives of the Ricci tensor in the definition \eqref{Cotton_tensor} and \eqref{Cotton-York_tensor} implies that the field equations \eqref{TMG_field_eqs} are partial differential equations of the third order. Standard Einstein's field equations in 3$D$ are recovered from \eqref{TMG_field_eqs} in the limit ${\mu \rightarrow \infty }$ and ${\sigma = 1}$.

Alternatively, by taking the trace of the field equations \eqref{TMG_field_eqs} (recall that $Y_{ab}$ is traceless), we can express the Ricci scalar simply as
\begin{equation} \label{TMG_eq_Trace}
\sigma R = 6\Lambda - 16\pi \, T \, ,
\end{equation}
where ${T \equiv g^{ab}\, T_{ab}}$ is the trace of the energy-momentum tensor. Substituting this expression for the Ricci scalar into the TMG field equations \eqref{TMG_field_eqs}, we obtain their equivalent formulation
\begin{equation} \label{TMG_field_eq_Ricci}
\sigma \, R_{ab} + \frac{1}{\mu}\, Y_{ab} = 2\Lambda \, g_{ab} + 8\pi \, (T_{ab} - T\, g_{ab}) \, .
\end{equation}
From this form of the field equations it is clear that in TMG the \emph{vacuum} solutions are not limited to only the (anti)-de Sitter spacetime (when ${\Lambda \neq 0}$) and the Minkowski spacetime (when ${\Lambda =0}$). The Ricci tensor is no longer proportional to just the metric tensor, there now also appears the Cotton--York tensor $Y_{ab}$. Consequently, the set of all possible vacuum solutions in TMG is more extensive. For their comprehensive review see chapters 16--19 of the monograph \cite{Garcia-Diaz:2017}.
\newpage

\section{Equation of geodesic deviation}

In metric theories of gravity, the gravitational field is primarily represented by a specific spacetime metric $g_{ab}$, which is determined by the distribution of matter encoded in $T_{ab}$. This results in spacetime curvature, which influences the motion of particles. Such a curvature is best studied via the \emph{relative motion} between two neighbouring bodies. The relative motion of nearby freely falling test particles (moving along geodesics) is described by the equation of geodesic deviation. Derived already a century ago by Levi-Civita and Synge \cite{Levi-Civita:1926, Synge:1926, Synge:1934} for test particles without spin or charge in any spacetime dimension, it has the coordinate form
\begin{equation} \label{Geodesic_Deviation_eq}
\frac{\DD^2 Z^\mu}{\dd \tau ^2} = R{^\mu}_{\alpha \beta \nu}\, u^\alpha u^\beta Z^\nu \, ,
\end{equation}
where $Z^\mu$ is the separation vector, $R{^\mu}_{\alpha \beta \nu}$ is the Riemann curvature tensor, and $u^\alpha $ is the velocity of the reference test particle whose proper time is $\tau $. At a given time, the separation vector $\boldsymbol{Z}$ specifies the \emph{position} of the second test particle, in relation to the reference test particle. Its coordinates are given by an exponential map ${x^\mu _{\mathcal{Q}} = \exp _\mathcal{P} \, (\boldsymbol{Z})\, x^\mu _{\mathcal{P}}}$, where $\mathcal{P}$ denotes a point on the reference geodesic, while $\mathcal{Q}$ denotes a point on the neighbouring geodesic. The equation of geodesic deviation \eqref{Geodesic_Deviation_eq} describes their \emph{relative acceleration}, which is represented by the second absolute derivative of the separation vector $\boldsymbol{Z}$, defined in components as
\begin{equation} \label{absolute_derivative_of_Z}
\frac{\DD^2 Z^\mu}{\dd \tau ^2} \equiv \nabla _{\boldu} (\nabla _{\boldu}\, Z^\mu)=Z{^\mu}_{;\rho \lambda}\, u^\rho u^\lambda \, ,
\end{equation}
where $\nabla _{\boldu}$ denotes the covariant derivative in the velocity direction $\boldu$. The velocity is, by definition, tangent to the reference geodesic $\gamma (\tau )$ (that is ${u^\alpha \equiv \frac{\dd x^\alpha}{\dd \tau}}$), and as a result, it is parallely propagated, ${\nabla _{\boldu} \, u^\alpha =0}$. The second equality in the definition \eqref{absolute_derivative_of_Z} follows from this identity. Additionally, we assume that the geodesic $\gamma (\tau )$ is a worldline of a \emph{timelike} particle, therefore ${\boldu \cdot \boldu =-1}$. For further details on the equation of geodesic deviation, we refer to the classic textbooks \cite{MTW:1973, Wald:1984, HEL:2006, dFB:2010}.

The equation of geodesic deviation in the form \eqref{Geodesic_Deviation_eq} is actually a \emph{linear approximation} in separation, and it is valid only for small relative speeds of neighbouring particles. Extensions to arbitrary relative velocities were explored in \cite{Hodgkinson:1972, Bazanski:1977, AP:1979, LN:1979, Ciufolini:1986, KHC:2001, CM:2002}, and elsewhere.

More importantly, this \emph{coordinate} form is not directly suitable for a proper physical interpretation of the relative motion of nearby particles. Correct physical analysis must be performed in relation to a given observer who carries out the relevant measurements. Such an approach was first employed by Pirani in standard 4$D$ General Relativity \cite{Pirani:1956, Pirani:1957}, and later generalized to higher dimensions \cite{PS:2012}. Using this formalism in 3$D$, an observer is naturally represented by a \emph{reference frame} $\{ \bolde _a \}$ consisting of a timelike vector corresponding to the observer's velocity (${\bolde _{(0)} = \boldu}$) and two spacelike vectors (${\bolde _{(1)},\, \bolde _{(2)}}$) spanning the orthogonal spatial directions. We thus assume that this triad satisfies the following normalization conditions
\begin{equation} \label{orthonormal_frame_condition}
\bolde _{a} \cdot \bolde _{b} \equiv g_{\alpha \beta}\, e^\alpha _{a}\, e^\beta _{b} = \eta _{ab} \, ,
\end{equation}
where ${\eta _{ab} \equiv \text{diag}(-1,\, 1,\, 1)}$ denotes the Minkowski metric tensor. Any vector can now be expressed in the observer's basis $\{ \bolde _a \}$, and in the case of the separation vector $\boldsymbol{Z}$ we get ${Z ^\mu = Z ^a\, e_a ^\mu}$, where $Z^a$ denote its components with respect to this basis. Such components can be obtained directly as projections of $\boldsymbol{Z}$ onto the dual basis 1-forms $ \{ \bolde ^a \}$, which by definition satisfy
\begin{equation} \label{dual_basis_condition}
e ^\alpha _{a} \, e _\alpha ^{b}  = \delta ^{b} _{a} \, .
\end{equation}
(We assign a dual 1-form $\bolde ^a$ corresponding to a basis vector $\bolde _a$ by the relation ${e _\alpha ^{a} \equiv \eta ^{ab}\, g _{\alpha \beta}\, e ^\beta _{b}}$.) The components of the separation vector $\boldsymbol{Z}$ can then be expressed as ${Z ^a = \boldsymbol{Z} \cdot \, \bolde ^a}$, which are the actual positions measured by an observer in its reference frame.

To determine the time evolution of these components $Z ^a (\tau)$, we project the geodesic deviation equation \eqref{Geodesic_Deviation_eq} onto the dual frame $\{ \bolde ^a \}$. For simplicity, we assume that the observer is co-moving with the reference test particle, that is ${\bolde _{(0)} = \boldu}$. In view of \eqref{dual_basis_condition} we get ${e ^{(0)} _\mu = -u _\mu }$, and the projection of \eqref{Geodesic_Deviation_eq} onto this 1-form reads
\begin{equation} \label{geodesic_deviation_0th_component}
\frac{\DD ^2 {Z}^{(0)}}{\dd \tau ^2} = -u_\mu \, R{^\mu}_{\alpha \beta \nu}\, u^\alpha u^\beta Z^\nu = 0\, ,
\end{equation}
due to the antisymmetries of the Riemann tensor. Since the velocity $\boldu$ is parallely propagated, we can include it inside the second absolute derivative. The equation \eqref{geodesic_deviation_0th_component} thus implies that the temporal separation $Z^{(0)}(\tau )$ of neighbouring geodesic particles is \emph{at most linear} in $\tau$. We further assume that the times of the reference particle and the nearby particle are synchronized, so that by a suitable choice of such initial conditions we get
\begin{equation}
Z^{(0)} = 0 \, .
\end{equation}
The particles remain synchronized throughout the time evolution, and they remain on the same ${\tau = \text{const.}}$ hypersurfaces.

Therefore, the dynamics of the system of nearby test particles is encoded only in the \emph{two spatial directions} $\bolde _{(1)},\, \bolde _{(2)}$ of the observer. Such a time evolution is described by projecting the equation of geodesic deviation \eqref{Geodesic_Deviation_eq} onto this subspace. It has the form
\begin{equation} \label{Geodesic_Deviation_eq_frame}
\ddot{Z}^{(i)} = R{^{(i)}}_{(0)(0)(j)}\, Z^{(j)} \, ,
\end{equation}
for ${i,\, j = 1,\, 2}$, where the spatial projection of the relative acceleration is ${\ddot{Z}^{(i)} \equiv {e^{(i)}}_{\mu}\, \dfrac{\DD ^2 Z^\mu}{\dd \tau ^2}}$, and ${R{^{(i)}}_{(0)(0)(j)} \equiv {e^{(i)}}_{\mu} \, R{^\mu}_{\alpha \beta \nu}\, u^\alpha u^\beta\, {e^\nu }_{(j)}}$ are the relevant components of the Riemann tensor. Let us emphasize that, in general, $\ddot{Z}^{(i)}$ \emph{is not equal} to the second absolute derivative $\dfrac{\DD ^2 Z^{(i)}}{\dd \tau ^2}$ of the separation vector frame components $Z^{(i)}$. They are equivalent if and only if the observer's spatial frame is parallelly transported along the reference geodesic.

The key equation \eqref{Geodesic_Deviation_eq_frame} can be analyzed further by decomposing the Riemann tensor. In three spacetime dimensions, the traceless part of the Riemann curvature tensor, i.e. the Weyl tensor, is identically zero. Consequently, the curvature tensor is fully determined by the Ricci tensor $R_{ab}$ and the Ricci scalar $R$, and it can be written as
\begin{equation} \label{Riemann_tensor}
R_{abcd} = 2\, (g_{a[c}\, R_{d]b} - g_{b[c}\, R_{d]a} ) - R\, g_{a[c}g_{d]b} \, .
\end{equation}
Using the specific field equations of TMG \eqref{TMG_field_eq_Ricci}, we can express the Ricci tensor $R_{ab}$, and the Ricci scalar $R$ from \eqref{TMG_eq_Trace}. Substituting these into \eqref{Riemann_tensor} we get an explicit expression for the Riemann curvature tensor, namely
\begin{align}
\sigma \, R_{abcd} = 2&\Lambda \, g_{a[c}g_{d]b} -\frac{2}{\mu}\, (g_{a[c}\, Y_{d]b} - g_{b[c}\, Y_{d]a} ) \label{Riemann_tensor_Cotton-York} \\
&+ 16\pi \, (g_{a[c}\, T_{d]b} - g_{b[c}\, T_{d]a} ) - 16\pi \, T \, g_{a[c}g_{d]b} \, . \nonumber
\end{align}
In TMG, the curvature of vacuum spacetimes is thus entirely described by the cosmological constant $\Lambda $, and by the Cotton--York tensor $Y_{ab}$. For non-vacuum spacetimes, additional terms are involved representing matter fields contained in $T_{ab}$ and its trace $T$.

The relevant triad components of the Riemann tensor, which appear in equation \eqref{Geodesic_Deviation_eq_frame}, can now be evaluated using equation \eqref{Riemann_tensor_Cotton-York}. This leads to the following expressions
\begin{align}
\sigma \, R{^{(1)}}_{(0)(0)(1)} &= \Lambda + \frac{1}{\mu}\, \big( Y_{(0)(0)}-Y_{(1)(1)}\big) + 8\pi \, \big( T_{(1)(1)}-T_{(0)(0)}-T\big) \, ,\nonumber\\
\sigma \, R{^{(1)}}_{(0)(0)(2)} &= -\frac{1}{\mu}\, Y_{(1)(2)} + 8\pi \, T_{(1)(2)} \, ,\label{Riemann_frame_components}\\
\sigma \, R{^{(2)}}_{(0)(0)(2)} &= \Lambda + \frac{1}{\mu}\, \big( Y_{(0)(0)}-Y_{(2)(2)}\big) + 8\pi \, \big( T_{(2)(2)}-T_{(0)(0)}-T\big) \, . \nonumber
\end{align}
Recall also that ${R{^{(i)}}_{(0)(0)(j)}=R_{(i)(0)(0)(j)}}$, which is a consequence of \eqref{dual_basis_condition}, as the spatial part of the Minkowski metric is just an identity. The triad components of the energy-momentum tensor are ${T_{(i)(j)} \equiv T_{\alpha \beta} \, e^\alpha _{(i)}e^\beta _{(j)}}$, while ${Y_{(i)(j)} \equiv Y_{\alpha \beta} \, e^\alpha _{(i)}e^\beta _{(j)}}$ represent the components of the Cotton--York tensor.

In our recent work \cite{PP:2023, PP:2024} we introduced a new convenient method for investigation of an algebraic structure of spacetimes in 3$D$. In this method the Cotton tensor \eqref{Cotton_tensor} is the key object. It proved to be useful to represent the information contained in its five independent components as real scalar quantities $\Psi _A$, which we call \emph{the Cotton scalars}. They are defined as specific projections of the Cotton tensor $C_{abc}$ onto a null triad ${\{ \boldk,\, \boldl,\, \boldm \}}$ satisfying the normalization conditions
\begin{align} \label{null_frame_conditions}
\setlength{\arraycolsep}{2pt}
\boldk \cdot \boldk &= 0 = \boldl \cdot \boldl \, , \hspace{-3em} & \hspace{-3em} \boldk \cdot \boldl &= -1 \, ,\nonumber \\
\boldm \cdot \boldk &= 0 = \boldm \cdot \boldl \, , \hspace{-3em} & \hspace{-3em} \boldm \cdot \boldm &=1 \, .
\end{align}
Actually, these five real Cotton scalars $\Psi _A$ are the lower-dimensional analogues of the famous Newman--Penrose scalars constructed using the Weyl tensor, see (3.59) in \cite{SKMHH:2003} or (2.8) in \cite{GP:2009}. In 3$D$, the Cotton scalars were introduced in equation (2.3) of \cite{PP:2023} (and in equation (9) of \cite{PP:2024}), namely
\begin{align}
\Psi _0 &\equiv C_{abc}\, k^a m^b k^c \, ,\nonumber \\
\Psi _1 &\equiv C_{abc}\, k^a l^b k^c \, ,\nonumber \\
\Psi _2 &\equiv C_{abc}\, k^a m^b l^c \, ,\label{Psi_A}\\
\Psi _3 &\equiv C_{abc}\, l^a k^b l^c \, ,\nonumber \\
\Psi _4 &\equiv C_{abc}\, l^a m^b l^c \, . \nonumber
\end{align}
Given the relation \eqref{Cotton-York_tensor} between the Cotton tensor $C_{abc}$ and the Cotton--York tensor $Y_{ab}$, we can express the components of the the Cotton--York tensor in terms of these Cotton scalars $\Psi _A$. As shown in equation (72) of \cite{PP:2024}, the corresponding null-frame projections of $Y_{ab}$ are
\begin{align}
Y_{ab}\, k^a k^b&=-\Psi _0 \, , & Y_{ab}\, k^a l^b&=-\Psi _2 \, , & Y_{ab}\, k^a m^b&=-\Psi _1 \, , \nonumber \\
Y_{ab}\, l^a l^b&=\Psi _4 \, , & Y_{ab}\, l^a m^b&=\Psi _3 \, , & Y_{ab}\, m^a m^b&=-2 \, \Psi _2 \, . \label{Cotton_York_null_projections}
\end{align}

These relations can be employed to express the triad components of the Cotton--York tensor which appear in equations \eqref{Riemann_frame_components} in terms of the Cotton scalars $\Psi _A$. An orthonormal frame $\{ \bolde _a \}$, defined in \eqref{orthonormal_frame_condition}, is related to the null basis ${\{ \boldk,\, \boldl,\, \boldm \}}$, which satisfies \eqref{null_frame_conditions}, by the following simple formulae,
\begin{equation} \label{orthonormal_basis_using_null_vectors}
\bolde _{0} = \frac{1}{\sqrt{2}}\, (\boldk + \boldl) \, ,\qquad \bolde _1 = \frac{1}{\sqrt{2}}\, (\boldk - \boldl ) \, ,\qquad \bolde _2 = \boldm \, .
\end{equation}
Inserting these relations into the definitions \eqref{Psi_A} of the frame components of the Cotton--York tensor, and applying \eqref{Cotton_York_null_projections}, we obtain
\begin{align}
Y_{(0)(0)} &= \tfrac{1}{2}(\Psi _4 - \Psi _0 - 2\Psi _2) \, , & Y_{(1)(1)} &= \tfrac{1}{2}(\Psi _4 - \Psi _0 + 2\Psi _2) \, ,\nonumber\\
Y_{(1)(2)} &= -\tfrac{1}{\sqrt{2}}(\Psi _1 + \Psi _3)\, , & Y_{(2)(2)} &= -2\Psi _2 \, .
\end{align}
The frame components $Y_{ab}$ are thus fully determined by a specific combination of the Cotton scalars $\Psi _A$. Based on these expressions, we can rewrite the curvature components \eqref{Riemann_frame_components} in the following way
\begin{align}
\sigma \, R{^{(1)}}_{(0)(0)(1)} &= \Lambda - \frac{2}{\mu}\, \Psi _2 + 8\pi \, \big( T_{(1)(1)}-T_{(0)(0)}-T\big) \, ,\nonumber \\
\sigma \, R{^{(1)}}_{(0)(0)(2)} &= \frac{1}{\sqrt{2}\, \mu}\, \big( \Psi _1 + \Psi _3\big) + 8\pi \, T_{(1)(2)} \, , \label{Riemann_components_Cotton_scalars}\\
\sigma \, R{^{(2)}}_{(0)(0)(2)} &= \Lambda - \frac{1}{2\mu }\, \big( \Psi _0 - \Psi _4 - 2\, \Psi _2\big) + 8\pi \, \big( T_{(2)(2)}-T_{(0)(0)}-T\big) \, . \nonumber
\end{align}
The Cotton scalars $\Psi _A$ now fully describe the contributions arising from the vacuum gravitational field equations of TMG to curvature (which was previously represented in \eqref{Riemann_frame_components} by the Cotton--York tensor $Y_{ab}$). This implies that the equation of geodesic deviation \eqref{Geodesic_Deviation_eq_frame} can finally be formulated using \eqref{Riemann_components_Cotton_scalars} in an \emph{invariant explicit form} as
\begin{align}
\sigma \, \ddot{Z}^{(1)} &= \Lambda \, Z^{(1)} - \frac{2}{\mu}\, \Psi _2 \, Z^{(1)} + \frac{1}{\sqrt{2}\, \mu}\, \big( \Psi _1 + \Psi _3\big) \, Z^{(2)} \nonumber \label{geodesic_deviation_frame_1}\\
&\qquad + 8\pi \, \Big[ \big( T_{(1)(1)}-T_{(0)(0)}-T\big) \, Z^{(1)} + T_{(1)(2)}\, Z^{(2)} \Big] \, ,\\
\sigma \, \ddot{Z}^{(2)} &= \Lambda \, Z^{(2)} + \frac{1}{\sqrt{2}\, \mu}\, \big( \Psi _1 + \Psi _3\big) \, Z^{(1)} - \frac{1}{2\mu}\, \big( \Psi _0 - \Psi _4 - 2\Psi _2\big) \, Z^{(2)} \nonumber\\
&\qquad + 8\pi \Big[ \big( T_{(2)(2)}-T_{(0)(0)}-T\big) \, Z^{(2)} + T_{(1)(2)}\, Z^{(1)} \Big] \, . \label{geodesic_deviation_frame_2}
\end{align}
These two equations completely describe the influence of the gravitational field on the spatial relative motion of nearby test particles in any TMG 2+1 spacetime. The motion is fully determined by the cosmological constant $\Lambda $, and by the five Cotton scalars $\Psi _A$ defined in \eqref{Psi_A}. In the presence of matter, there are additional terms involving the frame components of the energy-momentum tensor $T_{ab}$. These equations will play a key role in the physical interpretation of the Cotton scalars $\Psi _A$, as will be discussed in detail in the next section.

\section{Effect of various components of the gravitational field on test particles}

To analyse the physical meaning of the individual Cotton scalars $\Psi _A$ in 3$D$ TMG, we will now investigate the motion of test particles in a \emph{vacuum}. This separates the effects of gravity from that of a matter. The equations \eqref{geodesic_deviation_frame_1} and \eqref{geodesic_deviation_frame_2} describing the relative acceleration in the absence of matter, ${T_{ab}=0}$, simplify to
\begin{align}
\sigma \, \ddot{Z}^{(1)} &= \Lambda \, Z^{(1)} - \frac{2}{\mu}\, \Psi _2 \, Z^{(1)} + \frac{1}{\sqrt{2}\, \mu}\, \big( \Psi _1 + \Psi _3\big) \, Z^{(2)} \, , \label{Vacuum_geodesic_deviation_1}\\
\sigma \, \ddot{Z}^{(2)} &= \Lambda \, Z^{(2)} + \frac{1}{\sqrt{2}\, \mu}\, \big( \Psi _1 + \Psi _3\big) \, Z^{(1)} - \frac{1}{2\mu }\, \big( \Psi _0 - \Psi _4 -2 \Psi _2 \big) \, Z^{(2)} \, .\label{Vacuum_geodesic_deviation_2}
\end{align}
First, notice that in the General Relativity limit, which is ${\mu \rightarrow \infty }$ and ${ \sigma =1}$, the gravitational field components are completely suppressed, ${ \ddot{Z}^{(i)}=\Lambda \, Z ^{(i)}}$ for ${i=1,\, 2 }$. The motion of test particles is then fully determined by the cosmological constant $\Lambda $.

Generally, in TMG we can categorize the gravitational influence as follows.
\begin{itemize}

\item{ {\bf The isotropic influence of the cosmological constant $\Lambda $:}

If only the cosmological constant $\Lambda $ is present (i.e., in conformally flat spacetimes) the equation of geodesic deviation simplifies considerably to a compact form
\begin{equation} \label{Influence_of_Lambda}
\sigma  \begin{pmatrix}
			\ddot{Z}^{(1)} \\
			\ddot{Z}^{(2)}
		  \end{pmatrix}  		= \Lambda \begin{pmatrix}
		  									1 & 0\\
		  									0 & 1
		  								  \end{pmatrix} \!  \begin{pmatrix}
		  								  					Z^{(1)}\\
		  								  					Z^{(2)}
		  								  				\end{pmatrix} \, .
\end{equation}
According to the sign of the cosmological constant, the effect can either be the \emph{isotropic} \emph{receding} of the test particles for ${\Lambda >0}$, or their isotropic \emph{focusing} for ${\Lambda <0}$. In the case ${\Lambda =0}$ the test particles are in inertial motion. Analogous result was obtained long ago by Synge \cite{Synge:1934} in standard 4$D$ relativity, for higher dimensional generalization see \cite{PS:2012}. Actually, the results obtained in \cite{PS:2012} correctly reduce also to lower (2+1) dimensions by putting ${D=3}$ therein. The character of the cosmological constant flips when we consider the opposite sign of the coupling constant (${\sigma =-1}$).
}

\item{ {\bf The transverse effect represented by $\Psi _0$:}

The equations \eqref{Vacuum_geodesic_deviation_1} and \eqref{Vacuum_geodesic_deviation_2} describing the relative deviation of test particles now reduce to
\begin{equation} \label{Influence_of_Psi_0}
\sigma  \begin{pmatrix}
			\ddot{Z}^{(1)} \\
			\ddot{Z}^{(2)}
		  \end{pmatrix}  		= -\frac{\Psi _0}{2\, \mu } \begin{pmatrix}
		  									0 & 0\\
		  									0 & 1
		  								  \end{pmatrix} \!  \begin{pmatrix}
		  								  					Z^{(1)}\\
		  								  					Z^{(2)}
		  								  				\end{pmatrix} \, .
\end{equation}
Since there is no acceleration in the $\bolde _{(1)}$ direction, the test particles are affected only in the perpendicular spatial direction $\bolde _{(2)}$. The scalar $\Psi _0$ thus represents the \emph{transverse} component of the gravitational field. Such an effect is analogous to gravitational waves in 4$D$ and in higher dimensions. Therefore, we may reasonably interpret this component as a \emph{gravitational wave} in TMG which propagates along the null direction $\boldl$ (i.e., the spatial direction $-\bolde _{(1)}$). The amplitude of the gravitational wave is damped by the mass parameter $\mu $ (for bigger $\mu$ the effect is smaller). Because the gravitational wave is given by a single real Cotton scalar $\Psi _0$, it has only a \emph{single polarization mode}. This is unlike in higher-dimensional and 4$D$ Einstein gravity where there are $\tfrac{1}{2}D(D-3)$ polarization modes of the \emph{Weyl} tensor, see \cite{PS:2012}. This identically vanishes in ${D=3}$, but is replaced by an analogous effect represented by the single $\Psi _0$ component of the Cotton tensor.
}

\item{{\bf The longitudinal effect represented by $\Psi _1 $:}

The influence of this component of the gravitational field combines the motion in \emph{both} spatial directions $\bolde _{(1)}$ and $\bolde _{(2)}$ according to the symmetric relation
\begin{equation} \label{Influence_of_Psi_1}
\sigma  \begin{pmatrix}
			\ddot{Z}^{(1)} \\
			\ddot{Z}^{(2)}
		  \end{pmatrix}  		= \frac{\Psi _1}{\sqrt{2}\, \mu } \begin{pmatrix}
		  									0 & 1\\
		  									1 & 0
		  								  \end{pmatrix} \! \begin{pmatrix}
		  								  					Z^{(1)}\\
		  								  					Z^{(2)}
		  								  				\end{pmatrix} \, .
\end{equation}
The acceleration in one spatial direction is proportional to the displacement of the test particle in the complementary direction. In view of the relations \eqref{orthonormal_basis_using_null_vectors} we have ${k_{(1)} \equiv \boldk \cdot \bolde _{(1)} >0 }$ and ${l_{(1)} \equiv \boldl \cdot \bolde _{(1)} <0}$. The longitudinal disturbance of the gravitational field $\Psi _1$ is thus oriented along the spatial direction $-\bolde _{(1)}$. Analogous effects are present in standard General Relativity \cite{Synge:1934} and also in higher dimensions \cite{PS:2012} for spacetimes of algebraic type III (or more general).
}

\item{ {\bf The Newton-type effect represented by $\Psi _2$:}

The relative motion of test particles is now given by
\begin{equation} \label{Interpret_eq_Psi_2}
\sigma  \begin{pmatrix}
			\ddot{Z}^{(1)} \\
			\ddot{Z}^{(2)}
		  \end{pmatrix}  		= \frac{\Psi _2}{\mu } \begin{pmatrix}
		  									-2 & 0\\
		  									0 & 1
		  								  \end{pmatrix} \! \begin{pmatrix}
		  								  					Z^{(1)}\\
		  								  					Z^{(2)}
		  								  				\end{pmatrix} \, .
\end{equation}
Interestingly, the influence of this Cotton component is similar to the tidal Weyl effects present in 4$D$ gravity. However, the rate of contraction in the spatial $\bolde _{(1)}$ direction is twice as big as expansion in the direction $\bolde _{(2)}$. Therefore, this effect does not preserve the volume spanned by these spatial basis vectors. This is different from standard General Relativity and higher dimensional gravity, where the matrix describing the geodesic deviation is traceless, see \cite{PS:2012}. The Cotton scalar $\Psi _2$ is typical for spacetimes of algebraic type D \cite{PP:2023, PP:2024}, where it is the only non-vanishing component of the Cotton tensor in a canonical frame.
}

\item{{\bf The longitudinal effect represented by $\Psi _3 $:}

The equation governing the influence of this component of the gravitational field is equivalent to that of the scalar $\Psi _1$, and is given by
\begin{equation} \label{Influence_of_Psi_3}
\sigma  \begin{pmatrix}
			\ddot{Z}^{(1)} \\
			\ddot{Z}^{(2)}
		  \end{pmatrix}  		= \frac{\Psi _3}{\sqrt{2}\, \mu } \begin{pmatrix}
		  									0 & 1\\
		  									1 & 0
		  								  \end{pmatrix} \! \begin{pmatrix}
		  								  					Z^{(1)}\\
		  								  					Z^{(2)}
		  								  				\end{pmatrix} \, .
\end{equation}
This is due to the symmetry in the definition of the $\Psi _1$ and $\Psi _3$ scalars \eqref{Psi_A}, which are invariant under the exchange of the null vectors ${\boldk \leftrightarrow \boldl}$. Therefore, the component $\Psi _3$ represents a longitudinal effect propagating with respect to the spatial direction $\bolde _1$, that is opposite to the spatial effect induced by $\Psi _1$.
}

\item{{\bf The transverse effect represented by $\Psi _4$:}

The influence on the motion of test particles by this Cotton scalar is described by
\begin{equation} \label{Influence_of_Psi_4}
\sigma  \begin{pmatrix}
			\ddot{Z}^{(1)} \\
			\ddot{Z}^{(2)}
		  \end{pmatrix}  		= \frac{\Psi _4}{2\, \mu } \begin{pmatrix}
		  									0 & 0\\
		  									0 & 1
		  								  \end{pmatrix} \! \begin{pmatrix}
		  								  					Z^{(1)}\\
		  								  					Z^{(2)}
		  								  				\end{pmatrix} \, .
\end{equation}
This equation is equivalent to the relation \eqref{Influence_of_Psi_0} given by the complementary $\Psi _0$ scalar because the definitions of $\Psi _0$ and $\Psi _4$, given by \eqref{Psi_A}, are invariant with respect to the interchange ${\boldk \leftrightarrow \boldl}$. This means that $\Psi _4$ represents a \emph{gravitational wave propagating along the null direction} $\boldk$ (i.e., in the spatial direction $\bolde _{(1)}$). We can naturally assume that the null basis has usual orientation of $\boldk$  being an outgoing null vector, while $\boldl$ is incoming. Therefore, we can interpret the scalar $\Psi _0$ as an amplitude of the incoming gravitational wave, while the scalar $\Psi _4$ represents an outgoing one. The phase of the outgoing gravitational wave is shifted from that of the incoming wave, because in this case the equation of geodesic deviation also differs by a minus sign. According to the algebraic classification \cite{PP:2023, PP:2024}, spacetimes of algebraic type~N have the only non-vanishing component $\Psi _4$ of the Cotton tensor. We therefore conclude that spacetimes of this type represent exact gravitational waves in three dimensions in TMG.
}

\end{itemize}

\section{Interpreting spacetimes of a specific algebraic type}
The formulation of the geodesic deviation equation in terms of its frame components \eqref{geodesic_deviation_frame_1}, \eqref{geodesic_deviation_frame_2} provides an invariant description, independent of the choice of coordinates. However, the basis ${\bolde _{(1)} ,\, \bolde _{(2)}}$ representing the spatial directions orthogonal to the velocity $\boldu $ of an observer, is ambiguous. As a result, the decomposition of the geodesic deviation is not unique because it depends on the choice of this basis, and also on $\boldu$. In this section, we propose a ``canonical'' \emph{interpretation frame} for any spacetime, which is based on its \emph{algebraic type}.
\vspace{2mm}

In \cite{PP:2023, PP:2024} we proved that any 2+1 geometry admits a special null vector~$\boldk$, called the \emph{Cotton-aligned null direction} (CAND), which is aligned with the eigenvector of the Cotton tensor $C_{abc}$. In the associated principal null frame ${\{\boldk ,\, \boldl ,\, \boldm \}}$, the Cotton scalars $\Psi _A$ \eqref{Psi_A} take a canonical form as shown in Table 1 (which repeats Table I in \cite{PP:2023}, and Table 1 in \cite{PP:2024}).
\begin{table}[!h]
\begin{center}
\begin{tabular}{ c c l l}
\hline
\\[-20pt]
Algebraic type && \qquad\quad Canonical values of $\Psi _A$\\[2pt]
\hline
\\[-8pt]
I   && ${\Psi _0=0}$                              & $\Psi _1\ne0$ \\[2pt]
II  && ${\Psi _0=\Psi _1=0}$                      & $\Psi _2\ne0$ \\[2pt]
III && ${\Psi _0=\Psi _1=\Psi _2=0}$              & $\Psi _3\ne0$ \\[2pt]
N   && ${\Psi _0=\Psi _1=\Psi _2=\Psi _3=0}$      & $\Psi _4\ne0$ \\[2pt]
D   && ${\Psi _0=\Psi _1 = 0 = \Psi _3=\Psi _4}$  & $\Psi _2\ne0$ \\[2pt]
O   && all ${\Psi_{\rm A}=0}$  & \\[2pt]
\hline
\end{tabular}
\caption{The algebraic types of 2+1 geometries. These are identified by the canonical values of the Cotton scalars $\Psi _A$.}
\label{Tab-classification}
\end{center}
\end{table}

\newpage
Now, we naturally require that the observer's interpretation frame ${ \{ \boldu ,\, \bolde ^{\text{interp}} _{(1)} ,\, \bolde ^{\text{interp}} _{(2)} \} }$ is chosen in such a way that the corresponding null triad ${\{\boldk ^{\text{interp}},\, \boldl ^{\text{interp}},\, \boldm ^{\text{interp}}\}}$, given by the relations analogous to \eqref{orthonormal_basis_using_null_vectors},
\begin{equation} \label{null_interpretation_frame}
\boldk ^{\text{interp}}= \frac{1}{\sqrt{2}}(\boldu + \bolde _{(1)} ^{\text{interp}} )\, , \qquad \boldl ^{\text{interp}}= \frac{1}{\sqrt{2}}(\boldu - \bolde _{(1)} ^{\text{interp}} ) \, ,\qquad \boldm ^{\text{interp}}= \bolde _{(2)} ^{\text{interp}} \, ,
\end{equation}
is a principal null basis, that is $\boldk ^{\text{interp}}$ is proportional to a CAND $\boldk$. By choosing the specific CAND $\boldk $ in a spacetime of given algebraic type, we can construct the null vector $\boldk ^{\text{interp}}$, which due to \eqref{null_interpretation_frame} must satisfy ${\boldk ^{\text{interp}} \cdot \boldu = -1/\sqrt{2}}$, by a simple boost transformation
\begin{equation} \label{Boost}
\boldk ^{\text{interp}} = B \, \boldk \, ,
\end{equation}
where the boost parameter $B$ has the specific value
\begin{equation} \label{reciprocal_boost}
B = -\frac{1}{\sqrt{2}\, (\boldk \cdot \boldu )} \, .
\end{equation}
Such a parameter $B$ is well defined because ${\boldk \cdot \boldu \neq 0}$ for an arbitrary timelike velocity vector $\boldu$. Moreover, the relations \eqref{null_interpretation_frame} determine the spatial interpretation basis vector $\bolde ^{\text{interp}}_{(1)}$, which can be expressed as ${\bolde _{(1)} ^{\text{interp}} \equiv \sqrt{2}\, \boldk ^{\text{interp}} - \boldu}$. Substituting this form of the vector $\bolde _{(1)} ^{\text{interp}}$ into the expression for $\boldl ^{\text{interp}}$ in \eqref{null_interpretation_frame}, we find that it has the form
\begin{equation} \label{Interpretation_null_vector}
\boldl ^{\text{interp}}= \sqrt{2} \, \boldu - \boldk ^{\text{interp}}\, .
\end{equation}
Actually, this condition is sufficient to uniquely determine the remaining vector $\boldm ^{\text{interp}}$. To satisfy the relation ${ \boldk ^{\text{interp}} \cdot \boldl ^{\text{interp}} = -1 }$, the null interpretation frame and the principal null frame must be related by a boost $B$, and a null rotation $L$ with fixed $\boldk ^{\text{interp}}$, which is given by (41) in \cite{PP:2024}, that is
\begin{equation} \label{null_rotation_fixed_k}
\boldk ^{\text{interp}}= B \, \boldk  \, ,\qquad \boldl ^{\text{interp}} = B^{-1} \, \boldl  + \sqrt{2}L \, \boldm + L^2 B\, \boldk  \, , \qquad \boldm ^{\text{interp}} = \boldm  + \sqrt{2}L \, B \, \boldk  \, .
\end{equation}
Therefore, by substituting the expression for $\boldl ^{\text{interp}}$ from \eqref{Interpretation_null_vector}, after explicitly expressing $\boldk ^{\text{interp}}$ from \eqref{Boost}, we get that the null rotation parameter $L$ satisfies the following quadratic equation
\begin{equation}
\sqrt{2} \, \boldu = B^{-1}\, \boldl + \sqrt{2}\, L\, \boldm  + (L^2+1)\, B \, \boldk \, .
\end{equation}
The parameter $L$ can now be easily determined by taking the scalar product of this equation with $\boldm $, which due to \eqref{null_frame_conditions} gives
\begin{equation} \label{rotation_parameter_L}
L = \boldm \cdot \boldu \, .
\end{equation}
The value of $L$ then uniquely determines the spatial vector $\bolde _{(2)} ^{\text{interp}}$ because ${ \boldm^{\text{interp}} = \bolde _{(2)} ^{\text{interp}} }$ in \eqref{null_rotation_fixed_k}. The principal null frame is equal to the interpretation frame (up to a boost) only when an observer is chosen in such a way that ${\boldm \cdot \boldu = 0}$.

Consequently, the Cotton scalars $\Psi _A ^{\text{interp}}$, expressed in the interpretation null triad \eqref{null_rotation_fixed_k}, are related to the canonical Cotton scalars $\Psi _A$, expressed in the principle null basis \eqref{Psi_A}, by the transformation
\begin{align}\label{Psi_null_rotation_fixed_k}
\Psi_0 ^{\text{interp}} &= B^2 \, \Psi _0 \, , \nonumber\\
\Psi_1 ^{\text{interp}} &= B \, \Psi _1 + \sqrt{2}\,L\, B^2 \, \Psi _0 \, , \nonumber\\
\Psi_2 ^{\text{interp}} &= \Psi _2 + \sqrt{2}\,L\, B \, \Psi _1 + L^2\, B^2 \, \Psi _0 \, ,\\
\Psi_3 ^{\text{interp}} &= B^{-1}\, \Psi _3 -3\sqrt{2}\,L\, \Psi _2 -3L^2\, B\, \Psi _1 - \sqrt{2}\,L^3\, B^2 \, \Psi _0 \, , \nonumber\\
\Psi_4 ^{\text{interp}} &= B^{-2}\, \Psi _4 +2\sqrt{2}\,L\, B^{-1}\, \Psi _3 -6L^2\, \Psi _2 -2\sqrt{2}\,L^3\, B\, \Psi _1-L^4\, B^2 \, \Psi _0 \,, \nonumber
\end{align}
see Eqs.~(43), (44) in \cite{PP:2024}. This local Lorentz transformation preserves the canonical form of the Cotton scalars for various algebraic types summarized in Table \ref{Tab-classification}. It means that the interpretation frame constructed in this way retains the canonical structure of the Cotton scalars. This allows us to invariantly interpret the spacetimes of specific algebraic type.

\begin{itemize}

\item {\bf Interpreting type I spacetimes:}

For type I spacetimes ${\Psi _0 = 0 = \Psi _0 ^{\text{interp}}}$, and the equations of geodesic deviation \eqref{Vacuum_geodesic_deviation_1} and \eqref{Vacuum_geodesic_deviation_2}, expressed in the interpretation frame $\bolde _{(1)} ^{\text{interp}}$, $\bolde _{(2)} ^{\text{interp}}$, has the form
\begin{align}
\sigma \, \ddot{Z}^{(1)} &= \Lambda \, Z^{(1)} - \frac{2}{\mu}\, \Psi ^{\text{interp}}_2 \, Z^{(1)} + \frac{1}{\sqrt{2}\, \mu}\big( \Psi ^{\text{interp}}_1 + \Psi ^{\text{interp}}_3\big) \, Z^{(2)} \, ,
\label{geodesic_deviation_type_I_(1)}\\
\sigma \, \ddot{Z}^{(2)} &= \Lambda \, Z^{(2)} + \frac{1}{\sqrt{2}\, \mu}\big( \Psi ^{\text{interp}}_1 + \Psi ^{\text{interp}}_3\big) \, Z^{(1)} + \frac{1}{2\mu }\big( \Psi ^{\text{interp}}_4 +2 \Psi ^{\text{interp}}_2 \big) \, Z^{(2)} \, .
\label{geodesic_deviation_type_I_(2)}
\end{align}
where the Cotton scalars are explicitly
\begin{align}
\Psi ^{\text{interp}}_1 &= B \, \Psi _1 \, , \nonumber \\
\Psi ^{\text{interp}}_2 &= \Psi _2 + \sqrt{2}\,L\, B \, \Psi _1 \, ,  \nonumber \\
\Psi ^{\text{interp}}_3 &= B^{-1}\, \Psi _3 -3\sqrt{2}L\, \Psi _2 - 3L^2\, B \, \Psi _1 \, ,  \\
\Psi ^{\text{interp}}_4 &= B^{-2}\, \Psi _4 + 2\sqrt{2}\, L\, B^{-1}\, \Psi _3 - 6L^2 \, \Psi _2 - 2\sqrt{2}L^3\, B \, \Psi _1 \, .  \nonumber
\end{align}
The gravitational influence on test particles in these spacetimes is quite general, however, there may exist special observers. For example, if a geodesic observer exists such that ${L = -\Psi _2/(\sqrt{2}\, B\, \Psi _1)}$ then ${\Psi _2 ^{\text{interp}} =0}$. On the other hand, if ${\Psi _1 ^{\text{interp}} = -\Psi _3 ^{\text{interp}}}$ then such an observer would only experience a Newton-like gravitational field $\Psi ^{\text{interp}}_2$, with the transverse effect given by $\Psi ^{\text{interp}}_4.$

\item {\bf Interpreting type II spacetimes:}

The condition for type II spacetimes in Table \ref{Tab-classification} implies ${\Psi _0 ^{\text{interp}} = \Psi _1 ^{\text{interp}} = 0}$. The equation of geodesic deviation \eqref{Vacuum_geodesic_deviation_1} and \eqref{Vacuum_geodesic_deviation_2} for these spacetimes then reads
\begin{align}
\sigma \, \ddot{Z}^{(1)} &= \Lambda \, Z^{(1)} - \frac{2}{\mu}\, \Psi ^{\text{interp}}_2 \, Z^{(1)} + \frac{1}{\sqrt{2}\, \mu}\, \Psi ^{\text{interp}}_3\, Z^{(2)} \, ,
\label{geodesic_deviation_type_II_(1)}\\
\sigma \, \ddot{Z}^{(2)} &= \Lambda \, Z^{(2)} + \frac{1}{\sqrt{2}\, \mu}\, \Psi ^{\text{interp}}_3\, Z^{(1)} + \frac{1}{2\mu }\big( \Psi ^{\text{interp}}_4 +2 \Psi ^{\text{interp}}_2 \big) \, Z^{(2)} \, ,
\label{geodesic_deviation_type_II_(2)}
\end{align}
with
\begin{align}
\Psi ^{\text{interp}}_2 &= \Psi _2 \, , \nonumber \\
\Psi ^{\text{interp}}_3 &= B^{-1}\, \Psi _3 -3\sqrt{2}L\, \Psi _2\, , \\
\Psi ^{\text{interp}}_4 &= B^{-2}\, \Psi _4 + 2\sqrt{2}\, L\, B^{-1} \, \Psi _3 - 6L^2 \Psi _2  \, . \nonumber
\end{align}
There may be special observes when ${L = \Psi _3/(3\sqrt{2}\, B \, \Psi _2)}$ which do not feel the tidal effects (${\Psi _3 ^{\text{interp}} = 0}$). In the special case when ${\Psi _4 ^{\text{interp}} = -2\Psi _2 ^{\text{interp}}}$ there is only a longitudinal effect present in the spatial direction $\bolde _{(2)} ^{\text{interp}}$. However, all type II spacetimes exhibit a Newton-like gravitational field given by $\Psi _2 ^{\text{interp}}$.

\item {\bf Interpreting type III spacetimes:}

According to Table \ref{Tab-classification}, ${\Psi _0 ^{\text{interp}} = \Psi _1 ^{\text{interp}} = \Psi _2 ^{\text{interp}} = 0}$, and the goedesic deviation \eqref{Vacuum_geodesic_deviation_1} and \eqref{Vacuum_geodesic_deviation_2} for this case simply is
\begin{align}
\sigma \, \ddot{Z}^{(1)} &= \Lambda \, Z^{(1)} + \frac{1}{\sqrt{2}\, \mu}\, \Psi ^{\text{interp}}_3\, Z^{(2)} \, ,
\label{geodesic_deviation_type_III_(1)}\\
\sigma \, \ddot{Z}^{(2)} &= \Lambda \, Z^{(2)} + \frac{1}{\sqrt{2}\, \mu}\, \Psi ^{\text{interp}}_3\, Z^{(1)} + \frac{1}{2\mu }\, \Psi ^{\text{interp}}_4 \, Z^{(2)} \, ,
\label{geodesic_deviation_type_III_(2)}
\end{align}
and the non-zero Cotton scalars are given by
\begin{align}
\Psi ^{\text{interp}}_3 &= B^{-1}\, \Psi _3 \, , \\
\Psi ^{\text{interp}}_4 &= B^{-2}\, \Psi _4 + 2\sqrt{2}\, L\, B^{-1}\, \Psi _3  \, .
\end{align}
These spacetimes always exhibit a longitudinal effect whose strength is determined purely by $\Psi _3 ^{\text{interp}}$. There may also exist unique observers which experience only this longitudinal influence of the gravitational field. They are determined by the null rotation parameter in the form ${L = -\Psi _4/(2\sqrt{2}\, B \, \Psi _3)}$, so that ${\Psi ^{\text{interp}}_4=0}$.
 
\item {\bf Interpreting type N spacetimes:}

The equations of geodesic deviation \eqref{Vacuum_geodesic_deviation_1} and \eqref{Vacuum_geodesic_deviation_2} simplify significantly in this case, since only the Cotton scalar $\Psi _4$ is non-vanishing, and they can be written as
\begin{align}
\sigma \, \ddot{Z}^{(1)} &= \Lambda \, Z^{(1)} \, , \label{geodesic_deviation_type_N_(1)}\\
\sigma \, \ddot{Z}^{(2)} &= \Lambda \, Z^{(2)} + \frac{\Psi ^{\text{interp}}_4}{2\mu }\, Z^{(2)} \, , \label{geodesic_deviation_type_N_(2)}
\end{align}
where
\begin{equation} \label{interp_Psi_4}
\Psi ^{\text{interp}}_4 = B^{-2} \, \Psi _4 \, .
\end{equation}
For any observer, the gravitational field of this algebraic type, exhibits a purely transverse effects on test particles. Therefore, we can generally conclude that these spacetimes represent gravitational waves in TMG. The motion of the observer only influences the strength of the gravitational wave amplitude, encoded in the parameter~$B$ in \eqref{reciprocal_boost}. It follows that the geodesic deviation in the transverse direction is given by
\begin{equation}
\sigma \, \ddot{Z}^{(2)} = \Lambda \, Z^{(2)} + \frac{(\boldk \cdot \boldu )^2}{\mu }\, \Psi _4 \, Z^{(2)} \, .
\end{equation}

\item {\bf Interpreting type D spacetimes:}

The geodesic deviation equations \eqref{Vacuum_geodesic_deviation_1} and \eqref{Vacuum_geodesic_deviation_2}, for type D spacetimes in the interpretation frame \eqref{null_rotation_fixed_k} are
\begin{align}
\sigma \, \ddot{Z}^{(1)} &= \Lambda \, Z^{(1)} - \frac{2}{\mu}\, \Psi ^{\text{interp}}_2 \, Z^{(1)} + \frac{1}{\sqrt{2} \, \mu}\, \Psi ^{\text{interp}}_3\, Z^{(2)} \, ,
\label{geodesic_deviation_type_D_(1)}\\
\sigma \, \ddot{Z}^{(2)} &= \Lambda \, Z^{(2)} + \frac{1}{\sqrt{2}\, \mu}\, \Psi ^{\text{interp}}_3\, Z^{(1)} + \frac{1}{2\mu } \, \big( \Psi ^{\text{interp}}_4 + 2\Psi ^{\text{interp}}_2 \big) \, Z^{(2)} \, , \label{geodesic_deviation_type_D_(2)}
\end{align}
with
\begin{align}
\Psi ^{\text{interp}}_2 &= \Psi _2 \, , \\
\Psi ^{\text{interp}}_3 &= -3\sqrt{2}\, L \, \Psi _2 \, ,\\
\Psi ^{\text{interp}}_4 &= -6\, L^2 \, \Psi _2 \, .
\end{align}
A general observer experiences the Newton-like component $\Psi ^{\text{interp}}_2$, as well as the longitudinal and transverse components $\Psi ^{\text{interp}}_3$ and $\Psi ^{\text{interp}}_4$ of the gravitational field. Interestingly, the strength of all of them is proportional to a \emph{single} non-vanishing Cotton scalar $\Psi _2$, and depends on the kinematics of the observer through the parameter $L$ given by \eqref{rotation_parameter_L}. The geodesic deviation equations can thus be written as
\begin{align}
\sigma \, \ddot{Z}^{(1)} &= \Lambda \, Z^{(1)} - \frac{2}{\mu}\, \Psi _2 \, Z^{(1)} - \frac{3}{\mu}\, (\boldm \cdot \boldu)\, \Psi _2\, Z^{(2)} \, ,\\
\sigma \, \ddot{Z}^{(2)} &= \Lambda \, Z^{(2)} - \frac{3}{\mu}\, (\boldm \cdot \boldu )\, \Psi _2\, Z^{(1)}
  + \frac{1}{\mu } \big[ 1 - 3\, (\boldm \cdot \boldu )^2 \big]  \Psi _2 \, Z^{(2)} \, .
\end{align}
For ${\boldm \cdot \boldu = 0}$ there is ${L=0}$, and the system simplifies considerably to \eqref{Influence_of_Lambda} with \eqref{Interpret_eq_Psi_2}.

\end{itemize}

We should also emphasize that the CANDs in 3$D$ can sometimes be complex. Fortunately, a purely complex CANDs exist only in two cases, namely in special $\text{I}_{\text{c}}$ and $\text{D}_{\text{c}}$ spacetimes. In these two cases there is no unique interpretation frame, otherwise we are always able to choose a \emph{real} CAND for the interpretation frame. For additional information see Section~13 of \cite{PP:2024}.

\section{Effect of matter on test particles}

The non-trivial coupling of matter to gravity in General Relativity and its modifications creates additional direct effects which influence the motion of test particles. They are encoded in the frame components of the energy-momentum tensor in the equation of geodesic deviation \eqref{geodesic_deviation_frame_1} and \eqref{geodesic_deviation_frame_2}. Focusing only on impact of such matter influence, the equations describing the relative acceleration of the test particles can be summarized as
\begin{align}
\sigma \, \ddot{Z}^{(1)} &= 8\pi \Big[ \big( T_{(1)(1)}-T_{(0)(0)}-T\big) \, Z^{(1)} + T_{(1)(2)}\, Z^{(2)} \Big] \, , \label{geodesic_deviation_for_matter_1}\\
\sigma \, \ddot{Z}^{(2)} &= 8\pi \Big[ \big( T_{(2)(2)}-T_{(0)(0)}-T\big) \, Z^{(2)} + T_{(1)(2)}\, Z^{(1)} \Big] \, . \label{geodesic_deviation_for_matter_2}
\end{align}
In a general situation, the gravitational influence could not be neglected, unless the spacetime is conformally flat ($\Psi _A$ vanish for all $A$). However, it will be illustrative to investigate these effects separately now. For that reason we will consider several important types of matter, namely pure radiation, perfect fluid, and electromagnetic field.

\begin{itemize}

\item{{\bf Pure radiation:}

Pure radiation, or a null dust, is fully described by a single parameter $\rho $ which is the density of radiation. The energy-momentum tensor of the null dust aligned along the null direction $\boldk$ reads
\begin{equation}
T_{ab} = \rho \, k_a k_b \, .
\end{equation}
The analysis is significantly simplified if we consider an observer whose basis is adapted to this privileged null direction in terms of the relations \eqref{orthonormal_basis_using_null_vectors}. The relevant frame components of the energy-momentum tensor can then be easily evaluated as
\begin{equation}
T_{(0)(0)} = T_{(1)(1)} = \tfrac{1}{2}\, \rho \, ,
\end{equation}
all the remaining frame components are zero, as well as the trace ${T=0}$. This reduces the equation of geodesic deviation significantly, and it can be compactly expressed as
\begin{equation}
\sigma  \begin{pmatrix}
			\ddot{Z}^{(1)} \\
			\ddot{Z}^{(2)}
		  \end{pmatrix}  		= -4\pi \rho \begin{pmatrix}
		  									0 & 0\\
		  									0 & 1
		  								  \end{pmatrix} \! \begin{pmatrix}
		  								  					Z^{(1)}\\
		  								  					Z^{(2)}
		  								  				\end{pmatrix} \, .
\end{equation}
The influence of the pure radiation is \emph{purely transversal}, as the deviation of geodesics happens only in the spatial direction $\bolde _{(2)}$. In fact, the effect is reminiscent of the influence of a gravitational wave represented by $\Psi _4$. By taking the density of the null dust as ${\rho = -\tfrac{1}{8\pi \mu}\Psi _4}$, the deviation would be described exactly as for such gravitational wave. Moreover, the corresponding equations derived in higher dimensions in \cite{PS:2012} reduce correctly to the ${D=3}$ case.
}

\item{{\bf Perfect fluid:}

For a perfect fluid moving along the timelike velocity vector field $\boldu$, and described by density $\rho $ and isotropic pressure $p$, the energy-momentum tensor is given by
\begin{equation}
T_{ab} = (\rho + p)\, u_a u_b + p\, g_{ab} \, .
\end{equation}
Assuming an observer co-moving with the fluid, we can take the velocity vector $\boldu$ of the fluid as the basis timelike vector $\bolde _{(0)}$ of the observer. Without loss of generality we can also consider the spatial basis to be the same for the fluid and for the observer. Projection of the energy-momentum tensor onto such a basis leads to
\begin{equation}
T_{(0)(0)} = \rho \, ,\qquad T_{(1)(1)}=T_{(2)(2)}=p\, ,\qquad T = -\rho +2p \, ,
\end{equation}
while all the remaining projections are zero. The relative acceleration of test particles \eqref{geodesic_deviation_for_matter_1}, \eqref{geodesic_deviation_for_matter_2} is then described by a simple expression
\begin{equation}
\sigma  \begin{pmatrix}
			\ddot{Z}^{(1)} \\
			\ddot{Z}^{(2)}
		  \end{pmatrix}  		= -8\pi \, p \begin{pmatrix}
		  									1 & 0\\
		  									0 & 1
		  								  \end{pmatrix} \! \begin{pmatrix}
		  								  					Z^{(1)}\\
		  								  					Z^{(2)}
		  								  				\end{pmatrix} \, .
\end{equation}
Interestingly, the value of the acceleration is completely independent of the density $\rho $ of the fluid. It only depends on the pressure $p$ which manifests itself as an isotropic influence. In fact, if the value of the pressure is ${p=-\tfrac{1}{8\pi}\Lambda }$, the effect is exactly the same as the effect of the cosmological constant. These results can again be obtained by putting ${D=3}$ in the general ${D \geq 4}$ expressions derived in \cite{PS:2012}. As a consequence, for \emph{dust} particles (${p=0}$) there is locally no deviation of nearby geodesics in three dimensions. This fact was already pointed out by Giddings, Abbott and Kucha\v{r} \cite{GAK:1984} and also later by Garc\'ia-D\'iaz in the monograph \cite{Garcia-Diaz:2017}.
}

\item{{\bf Electromagnetic field:}

The energy-momentum tensor of an electromagnetic field is defined by the expression
\begin{equation}
T_{ab} = \frac{1}{4\pi }\Big( F_{ac}\, F{_b}^{c}-\frac{1}{4}F_{cd}\, F^{cd} \Big) \, .
\end{equation}
This has a well-known property that only in 4$D$ the trace of such energy-momentum tensor vanishes. In this sense, 4$D$ is special because in higher dimensional gravity, as well as in 3$D$ gravity, the trace is non-trivial. This makes the equation of geodesic deviation more complicated. However, we can use the Newman--Penrose formalism to simplify the analysis. In three dimension, the scalars of the electromagnetic field (defined in \cite{PP:2022}) are
\begin{equation} \label{Newman-Penrose_el._mag_field}
\phi _0 \equiv F_{ab}\, k^a m^b \, ,\qquad \phi _1 \equiv F_{ab}\, k^a l^b \, ,\qquad \phi _2 \equiv F_{ab}\, m^a l^b \, .
\end{equation}
These encode the three independent components of the electromagnetic field as specific projections of the Maxwell tensor. We can express the frame components of the energy-momentum tensor, relating the orthonormal frame of the observer to a null basis using \eqref{orthonormal_basis_using_null_vectors}. The metric tensor, written in terms of the null basis vectors, is
\begin{equation} \label{metric_in_null_basis}
g_{ab} = -k_a \, l_b - l_a \, k_b + m_a \, m_b \, .
\end{equation}
Aligning the basis of the observer with the privileged null basis used in \eqref{Newman-Penrose_el._mag_field} and \eqref{metric_in_null_basis} leads to the following frame components of the energy-momentum tensor,
\begin{align}
4\pi \, T_{(0)(0)} &= \tfrac{1}{2}(\phi _0 ^2 + \phi _1 ^2 + \phi _2 ^2 ) \, ,\nonumber \\
4\pi \, T_{(1)(1)} &= \tfrac{1}{2}(\phi _0 ^2 - \phi _1 ^2 + \phi _2 ^2 ) \, ,\nonumber \\
4\pi \, T_{(2)(2)} &= \phi _0 \, \phi _2 + \tfrac{1}{2}\, \phi _1 ^2 \, ,\\
4\pi \, T_{(1)(2)} &= \tfrac{1}{\sqrt{2}}\, \phi _1 (\phi _0 - \phi _2 )\, ,\nonumber \\
4\pi \, T &= \, \phi _0 \, \phi _2 - \tfrac{1}{2}\, \phi _1 ^2\, .\nonumber
\end{align}
Miraculously, this leads to a big simplification of the equation of geodesic deviation \eqref{geodesic_deviation_for_matter_1} and \eqref{geodesic_deviation_for_matter_2}, namely
\begin{align}
\sigma \, \ddot{Z}^{(1)} &= -\big( \phi _1 ^2+2\, \phi _0 \, \phi _2\big) \, Z^{(1)} + \sqrt{2}\, \phi _1 \big( \phi _0 - \phi _2\big) \, Z^{(2)} \, ,\\
\sigma \, \ddot{Z}^{(2)} &=  \sqrt{2}\, \phi _1 \big( \phi _0 - \phi _2\big) \, Z^{(1)} + \big( \phi _1 ^2 - \phi _0 ^2 - \phi _2 ^2\big) \, Z^{(2)} \, .
\end{align}
The influence on the motion of test particles can now be discussed for specific components of the electromagnetic field, namely the null components $\phi _0$ and $\phi _2$ representing an electromagnetic wave, and the non-null component $\phi _1$.
\begin{itemize}

\item{{\bf The transverse influence of the null component $\phi _0$:}

Considering only the $\phi _0$ component of the electromagnetic field, the geodesic deviation equation simplifies significantly to
\begin{equation} \label{Influence_of_Phi_0}
\sigma  \begin{pmatrix}
			\ddot{Z}^{(1)} \\
			\ddot{Z}^{(2)}
		  \end{pmatrix}  		= -\phi _0 ^2 \begin{pmatrix}
		  									0 & 0\\
		  									0 & 1
		  								  \end{pmatrix} \! \begin{pmatrix}
		  								  					Z^{(1)}\\
		  								  					Z^{(2)}
		  								  				\end{pmatrix} \, .
\end{equation}
A non-trivial influence on the relative motion of test (uncharged) particles thus occurs only in the spatial direction $\bolde _{(2)}$. This is an analogous result for other forms of radiation, be it gravitational or pure radiation. In fact, a non-vanishing scalar $\phi _0$ represents an electromagnetic wave traveling along the $\boldl$ null direction. Also, since $\phi _0 ^2>0$ the effect can only be the focusing of nearby geodesics.
}

\item{{\bf The Coulomb-type influence of the non-null component $\phi _1$:}

For only $\phi _1$ non-vanishing, the equation of geodesic deviation is simply
\begin{equation} \label{Influence_of_Phi_1}
\sigma  \begin{pmatrix}
			\ddot{Z}^{(1)} \\
			\ddot{Z}^{(2)}
		  \end{pmatrix}  		= \phi _1 ^2 \begin{pmatrix}
		  									-1 & 0\\
		  									0 & 1
		  								  \end{pmatrix} \! \begin{pmatrix}
		  								  					Z^{(1)}\\
		  								  					Z^{(2)}
		  								  				\end{pmatrix} \, .
\end{equation}
The motion of the test particles is focused in the spatial direction $\bolde _{(1)}$ while in the complementary direction $\bolde _{(2)}$ the particles recede from each other. The effect is similar to the Newton-like influence of the $\Psi _2$ Cotton scalar of the gravitational field given by \eqref{Interpret_eq_Psi_2}. However, a major difference is that in the case of electromagnetic field the trajectories of the particles are deviated in such a way that there is \emph{no shear} present (meaning that the matrix describing the relative acceleration in \eqref{Influence_of_Phi_1} is traceless.).
}

\item{{\bf The transverse influence of the null component $\phi _2$:}

The influence of the remaining electromagnetic scalar $\phi _2$ is governed by an equivalent equation to that of $\phi _0$, \eqref{Influence_of_Phi_0}, which explicitly reads
\begin{equation} \label{Influence_of_Phi_2}
\sigma  \begin{pmatrix}
			\ddot{Z}^{(1)} \\
			\ddot{Z}^{(2)}
		  \end{pmatrix}  		= -\phi _2 ^2\begin{pmatrix}
		  									0 & 0\\
		  									0 & 1
		  								  \end{pmatrix} \! \begin{pmatrix}
		  								  					Z^{(1)}\\
		  								  					Z^{(2)}
		  								  				\end{pmatrix} \, .
\end{equation}
We thus obtain the usual effect of radiation as a transverse influence on the motion of test particles. This is not surprising, as the $\phi _2$ scalar represents an electromagnetic wave traveling along the $\boldk$ null direction. Notice again the symmetry in the definitions \eqref{Newman-Penrose_el._mag_field} which are, up to a sign, invariant with respect to interchange ${\boldk \leftrightarrow \boldl}$ of the null vectors. This explains the similarity of \eqref{Influence_of_Phi_0} and \eqref{Influence_of_Phi_2}.
}

\end{itemize}
}

\end{itemize}

\section{Explicit example: Type N wave metrics in TMG}

The field equations of TMG allow for a Brinkmann-like class of spacetimes characterized by a nonexpanding (twist and shear vanish identically in 3$D$ \cite{CPS:2010b,PSM:2019}), geodesic, and Killing null vector field $\boldk$. In the presence of the cosmological constant $\Lambda $, this vector field $\boldk$ is not covariantly constant, and these solutions are known as \emph{TN-waves} (see Chapter 18 in \cite{Garcia-Diaz:2017}). The general form of their line element, given by equations (18.2) and (18.8) in \cite{Garcia-Diaz:2017}, is\footnote{This metric is related to the one in (18.2) by a simple identification ${v \mapsto -r,\,\rho \mapsto x}$.}
\begin{equation} \label{TN-wave_metric}
\dd s^2 = \dd x^2 - 2\e ^{2\sqrt{-\Lambda }\, x}\, \dd u\, \dd r + H(u,x)\, \dd u^2 \, .
\end{equation}
These TN-wave spacetimes admit only zero and negative values of the cosmological constant ${\Lambda \leq 0}$, and the function $H(u,x)$ satisfies the \emph{vacuum field equation}~(18.9) in \cite{Garcia-Diaz:2017}, namely
\begin{equation} \label{TN-wave_field_eq}
H_{,xxx} - 3\sqrt{-\Lambda }\, H_{,xx} - 2\Lambda \, H_{,x} = \mu \, \big( 2\sqrt{-\Lambda }\, H_{,x} - H_{,xx}\big) \, .
\end{equation}
The choice of the coordinates \eqref{TN-wave_metric} is naturally adapted to the null vector field $\boldk$, which is expressed here as ${\boldk = \partial _r}$. Interestingly, this vector field is also the unique quadruple CAND present in type~N spacetimes. Using the relations \eqref{null_frame_conditions}, the corresponding null triad can be constructed as
\begin{equation} \label{TN-wave_CAND}
\boldk = \partial _r \, ,\qquad \boldl = \e ^{-2\sqrt{-\Lambda }\, x}\Big( \partial _u + \tfrac{1}{2} \e ^{-2\sqrt{-\Lambda }\, x}\, H \, \partial _r\Big) \, ,\qquad \boldm = \partial _x \, .
\end{equation}
The Cotton tensor \eqref{Cotton_tensor}, evaluated for the metric \eqref{TN-wave_metric}, has only a single independent component ${C_{xuu}= \tfrac{1}{2}(2\Lambda \, H_{,x}+3\sqrt{-\Lambda }\, H_{,xx}-H_{,xxx})}$, so that the corresponding Cotton scalars \eqref{Psi_A} in the null basis \eqref{TN-wave_CAND} are
\begin{align}
\Psi _0 &= \Psi _1 = \Psi _2 = \Psi _3 = 0 \, , \nonumber\\
\Psi _4 &= \tfrac{1}{2}\e ^{-4\sqrt{-\Lambda }\, x}\, \big( H_{,xxx} - 3\sqrt{-\Lambda}\, H_{,xx} - 2\Lambda \, H_{,x} \big) \, .
\end{align}
This is the canonical form of the Cotton scalars for type~N spacetimes, as given in Table~\ref{Tab-classification}, confirming that $\boldk$ is indeed the (quadruply degenerate) CAND. Using the field equation \eqref{TN-wave_field_eq} the key scalar $\Psi _4$ simplifies to
\begin{equation} \label{TMG vacuum}
\Psi _4 = \tfrac{1}{2}\, \mu \, \e ^{-4\sqrt{-\Lambda }\, x}\, \big( 2\sqrt{-\Lambda }\, H_{,x} - H_{,xx} \big) \, .
\end{equation}

Considering an arbitrary geodesic observer with the velocity ${\boldu = \dot{r}\, \partial _r + \dot{u}\, \partial _u + \dot{x}\, \partial _x}$, we get the boost parameter \eqref{reciprocal_boost} and the null rotation parameter \eqref{rotation_parameter_L} evaluated with respect to the principal null basis \eqref{TN-wave_CAND}, namely
\begin{equation}
B = \frac{\e ^{-2\sqrt{-\Lambda }\, x}}{\sqrt{2}\, \dot{u}} \, , \qquad L = \dot{x} \, .
\end{equation}
The interpretation null triad can thus be immediately constructed using \eqref{null_rotation_fixed_k} as
\begin{align} \label{TN-wave_interpretation_frame}
\boldk ^{\text{interp}} &= \frac{\e ^{-2\sqrt{-\Lambda }\, x}}{\sqrt{2}\, \dot{u}}\, \partial _r \, ,\nonumber \\
\boldl ^{\text{interp}} &= \Big(\sqrt{2}\,\dot{r}-\frac{\e^{-2\sqrt{-\Lambda }\,x}}{\sqrt{2}\,\dot{u}} \Big) \, \partial _r
+ \sqrt{2}\, \dot{u}\, \partial _u + \sqrt{2}\, \dot{x}\, \partial _x \, ,  \\
\boldm ^{\text{interp}} &= \frac{\dot{x}}{\dot{u}}\, \e ^{-2\sqrt{-\Lambda }\, x} \, \partial _r + \partial _x \, .\nonumber
\end{align}
The only non-trivial Cotton scalar transforms according to \eqref{interp_Psi_4} as ${\Psi _4 ^{\text{interp}} = B^{-2}\, \Psi _4}$, which for vacuum solutions \eqref{TMG vacuum} reduces to a compact expression
\begin{equation} \label{TN-wave_Psi_4}
\Psi _4 ^{\text{interp}} = \mu \, \dot{u}^2 \big( 2\sqrt{-\Lambda }\, H_{,x} - H_{,xx} \big) \, .
\end{equation}
Let us emphasize that the frame \eqref{TN-wave_interpretation_frame} is parallelly propagated only when ${\Lambda = 0}$. Even though a parallelly propagated frame is not necessary for physical interpretation of motion, its use simplifies the integration of the geodesic deviation.

Therefore, in the following we assume only the case with the \emph{vanishing cosmological constant} ${\Lambda = 0}$, in which case the line-element \eqref{TN-wave_metric} simplifies to
\begin{equation} \label{pp-wave_metric}
\dd s^2 = \dd x^2 - 2\, \dd u\, \dd r + H(u,x)\, \dd u^2 \, .
\end{equation}
This 2+1 spacetime metric represents \emph{pp-waves} in TMG because the (quadruply degenerate) null vector field $\boldk = \partial _r$ is now \emph{covariantly constant}. Moreover, the field equation \eqref{TN-wave_field_eq} for the function $H(u,x)$ can be fully integrated, see (18.32) and (18.33) in \cite{Garcia-Diaz:2017}. Its general solution is
\begin{equation} \label{pp-wave_metric_function}
H(u,x) = f_1(u)\, \e ^{-\mu \, x} + f_2 (u) \, x + f_3 (u) \, ,
\end{equation}
where $f_1$, $f_2$ and $f_3$ are arbitrary functions of the retarded time $u$. In fact, the functions $f_2$ and $f_3$ can be eliminated by a gauge transformation, and the only physical parameter is the \emph{profile $f_1(u)$ of the gravitational wave}, which can be prescribed \emph{arbitrarily}. This follows from the expression \eqref{TN-wave_Psi_4} for the scalar $\Psi _4^{\text{interp}}$ by substituting the metric function \eqref{pp-wave_metric_function} (and setting ${\Lambda = 0}$ therein). Indeed, we obtain
\begin{equation}\label{pp-wave_Psi_4_interp}
\Psi _4 ^{\text{interp}} = -\mu ^3\, \dot{u}^2\, f_1(u) \, \e ^{-\mu \, x}\,  .
\end{equation}
The corresponding interpretation frame \eqref{TN-wave_interpretation_frame}, for the present case ${\Lambda = 0}$, reduces to
\begin{equation} \label{pp-wave_interpretation_frame}
\boldk ^{\text{interp}} = \frac{1}{\sqrt{2}\, \dot{u}}\, \partial _r \, , \ \boldl ^{\text{interp}} = \sqrt{2}\Big( \dot{r} - \frac{1}{2\dot{u}} \Big) \partial _r + \sqrt{2}\, \dot{u}\, \partial _u + \sqrt{2}\, \dot{x}\, \partial _x \, , \ \boldm ^{\text{interp}} = \frac{\dot{x}}{\dot{u}} \, \partial _r + \partial _x \, .
\end{equation}
The equation of geodesic deviation for these type~N spacetimes is given by \eqref{geodesic_deviation_type_N_(1)}, \eqref{geodesic_deviation_type_N_(2)} and thus in the case of pp-waves it can be written as
\begin{align}
\ddot{Z}^{(1)} &= 0 \, , \nonumber\\
\ddot{Z}^{(2)} &= -\tfrac{1}{2}\, \mu ^2 \, \dot{u}^2 \, f_1(u) \, \e ^{-\mu \, x}\,Z^{(2)} \,  .  \label{pp-wave_geodesic_deviation}
\end{align}
The wave amplitude profile depends on the retarded coordinate $u$ via ${f_1(u)}$. This resembles pp-waves in 4$D$ and higher dimensions \cite{GP:2009, PS:2012}. In the transverse spatial direction, the wave-form is determined by an exponential function of $x$. It vanishes for ${x \to \infty}$, where the spacetime is flat (on the other hand, for ${x \to -\infty}$ there is a curvature singularity).

To explicitly solve the geodesic deviation equation \eqref{pp-wave_geodesic_deviation}, we  need to find the $\tau$-dependence of $u(\tau)$ and $x(\tau)$ along a timelike geodesic. Since  the Christoffel symbols for the metric \eqref{pp-wave_metric} are ${\Gamma^{u}_{\alpha \beta}=0}$, it follows that $\dot{u}$ remains \emph{constant} along any geodesic, so that
\begin{equation} \label{dot-u_0}
u(\tau) = \dot{u}_0 \, \tau + u_0 \,,
\end{equation}
where $\dot{u}_0,\, \tau _0$ are constants. The only non-vanishing Christoffel symbols that are relevant in the geodesic equation ${\tfrac{\dd ^2 x}{\dd \tau^2 } + \Gamma^x_{\alpha \beta} \, \tfrac{\dd x^\alpha}{\dd \tau }\, \tfrac{\dd x^\beta}{\dd \tau}=0}$  are
${\Gamma ^{x}_{uu} = \tfrac{1}{2} \big[ \mu \, f_1(u) \, \e ^{-\mu \, x} - f_2(u) \big]}$,
so that
\begin{equation} \label{pp-waves_geodesic_eq}
\frac{\dd ^2 x}{\dd \tau ^2} = \tfrac{1}{2} \big[ f_2(u) - \mu \, f_1(u) \, \e ^{-\mu \, x} \big] \, \dot{u}_0^2 \, .
\end{equation}
Therefore, to integrate this equation, we only need to prescribe the functions $f_1(u)$ and $f_2(u)$, in which $u(\tau)$ is just a linear function \eqref{dot-u_0} of the proper time $\tau$.

We now concentrate on the simplest case of \emph{particular vacuum} pp-wave solutions which, in the form of the metric \eqref{pp-wave_metric}, \eqref{pp-wave_metric_function}, have the following \emph{constant values} of the functions
\begin{equation}
f_1 (u) = f_1 = \text{const.} \, ,\qquad f_2(u)=0=f_3(u) \, .
\end{equation}
Examples of such solutions are the Percacci, Sodano and Vuorio pp-wave solution \cite{PSV:1987}, for which the constants are ${f_1 = - \mu ^2 / 4}$, ${f_2=0=f_3}$, and the Aragone pp-wave solution \cite{Aragone:1987}, for which ${f_1 = 4/\mu ^2}$, ${f_2=0=f_3}$, see Chapters 18.3.2 and 18.3.3 in \cite{Garcia-Diaz:2017}.

Depending on the \emph{sign} of the constant $f_1$, the geodesic equation \eqref{pp-waves_geodesic_eq} has two distinct solutions, namely
\begin{equation} \label{pp-wave_x_tau}
x(\tau ) = \begin{cases}
-\dfrac{1}{\mu}\, \log \Big[ \,\,\,\dfrac{1}{f_1}\Big( \dfrac{2\, c_0}{\mu \, \dot{u}_0} \Big) ^2\, \sec ^2 \big[ c_0 \, (\tau - \tau _0)\big]  \Big] & \text{for $f_1>0$\,,}\\[4mm]
-\dfrac{1}{\mu}\, \log \Big[ -\dfrac{1}{f_1}\Big( \dfrac{2\, c_0}{\mu \, \dot{u}_0} \Big) ^2\, \sech ^2 \big[ c_0 \, (\tau - \tau _0)\big]  \Big] & \text{for $f_1<0$\,,}
\end{cases}
\end{equation}
where $c_0$ and $\tau _0$ are arbitrary constants. 

The non-trivial geodesic deviation equation \eqref{pp-wave_geodesic_deviation} for \eqref{pp-wave_x_tau} becomes
\begin{equation} \label{pp-wave_geodesic_deviation_sol}
\dfrac{\dd ^2 Z^{(2)}}{\dd \tau ^2} = \begin{cases}
-2\, c_0^2 \, \sec ^2 \big[ c_0 \, (\tau - \tau _0)\big] \, Z^{(2)} & \text{for $f_1>0$\,,}\\[4mm]
\,\,2\, c_0^2 \, \sech ^2 \big[ c_0 \, (\tau - \tau _0)\big] \, Z^{(2)} & \text{for $f_1<0$\,.}
\end{cases}
\end{equation}
Notice that it is independent of $\mu$, and of the absolute value of $f_1$. Here we conveniently employed the fact that the interpretation frame is parallelly transported, and thus ${\ddot{Z}^{(2)}\equiv\dfrac{\DD ^2 Z^{(2)}}{\dd \tau ^2}=\dfrac{\dd ^2 Z^{(2)}}{\dd \tau ^2}}$.

\begin{figure}[t!]
\centering
\includegraphics[width=0.75\textwidth]{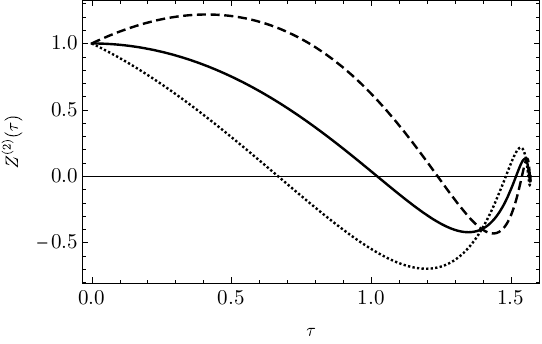}
\caption{Plot of the separation of geodesics in the transverse spatial direction $Z^{(2)}(\tau)$ given by \eqref{pp-wave_separation_vector_solutions} for ${f_1>0}$. The initial conditions for the geodesic equation were chosen such that ${c_0=1}$ and ${\tau_0 = 0}$. The function is evaluated for the initial condition ${Z^{(2)}(0)=1}$ and three values of the initial velocity, namely ${\dot{Z}^{(2)}(0) = +1, 0, -1}$ (dashed, solid, dotted curves, respectively). The domain of the plot ends at ${\tau = \pi /2}$ when the geodesic observers reach singularity at ${x=-\infty}$.}
\label{Plot_Z_f1>0}
\end{figure}

\begin{figure}[t!]
\centering
\includegraphics[width=0.75\textwidth]{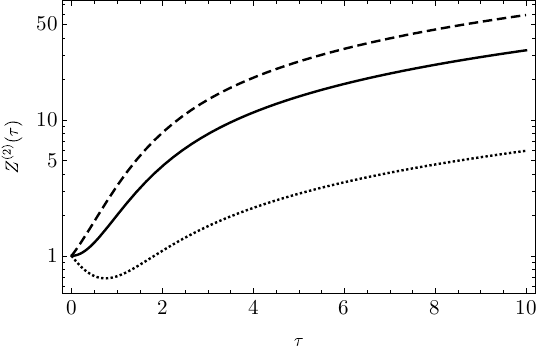}
\caption{Plot of the separation of geodesics in the transverse spatial direction $Z^{(2)}(\tau)$ given by \eqref{pp-wave_separation_vector_solutions} for ${f_1<0}$. The function is evaluated for the same initial conditions as in Figure~\ref{Plot_Z_f1>0}. Notice that the vertical axis has a logarithmic scale. The gravitational effect in this case is repulsive, and the geodesics never reach the singularity at ${x=-\infty}$.}
\label{Plot_Z_f1<0}
\end{figure}

This second-order differential equation is actually equivalent to the time-independent (real) Schr\"odinger equation with ${E=0}$ and the trigonometric \emph{P\"oschl--Teller potential} ${V(w) = - \sec ^2 w}$ for ${f_1>0}$, and the complementary hyperbolic P\"oschl--Teller potential ${V(w) = \sech ^2 w}$ for ${f_1<0}$ \cite{PT:1933}. Using a substitution ${z=\im \, \tan \big[ c_0 \, (\tau - \tau _0)\big]}$ in the case ${f_1>0}$, and ${z = \tanh \big[ c_0 \, (\tau - \tau _0)\big]}$ in the case ${f_1<0}$, the equation \eqref{pp-wave_geodesic_deviation_sol} is put into the form
\begin{equation} \label{pp-wave_geodesic_deviation_Legendre_form}
(1-z^2)\, \frac{\dd ^2 Z^{(2)}}{\dd z^2} - 2z\, \frac{\dd Z^{(2)}}{\dd z } - 2\, Z^{(2)} = 0 \, .
\end{equation}
This  is the \emph{Legendre equation} ${(1-\xi^2)\, y''-2\xi\, y' + \nu (\nu+1) \, y = 0}$, whose solutions $y(\xi)$ are linear combinations of the Legendre functions of the first kind $P_\nu (\xi)$ and the Legendre functions of the second kind $Q_\nu (\xi)$. In the case of equation \eqref{pp-wave_geodesic_deviation_Legendre_form} the parameter is
\begin{equation} \label{nu_Legendre_parameter}
\nu = -\tfrac{1}{2}\big(1 \pm \im \sqrt{7}\,\big),
\end{equation}
so that the general solutions are
\begin{equation} \label{pp-wave_separation_vector_solutions}
Z^{(2)}(\tau) = \begin{cases}
c_1 \, P_\nu \Big( \im \, \tan \big[ c_0 \, (\tau - \tau _0)\big] \Big) + c_2 \, Q_\nu \Big( \im \, \tan \big[ c_0 \, (\tau - \tau _0)\big] \Big)  & \text{for $f_1>0$\,,} \\[4mm]
c_1 \, P_\nu \Big( \tanh \big[ c_0 \, (\tau - \tau _0)\big] \Big) + c_2 \, Q_\nu \Big( \tanh \big[ c_0 \, (\tau - \tau _0)\big] \Big) & \text{for $f_1<0$\,,}
\end{cases}
\end{equation}
where $c_1$ and $c_2$ are arbitrary constants. Plot of this function determining the time evolution of the separation of the test particles for typical initial conditions in the case ${f_1>0}$ is given in Figure~\ref{Plot_Z_f1>0}. A complementary plot for the case $f_1<0$ is given in Figure~\ref{Plot_Z_f1<0}.

\section{Conclusions}

In this paper, we performed a systematic analysis of spacetimes in Topological Massive Gravity by studying the relative motion of uncharged, spinless test particles. We employed the equation of geodesic deviation, a well-established tool in both standard and higher-dimensional GR. It  proved to be valuable also for understanding exact solutions in this lower-dimensional modified theory of gravity. When expressed in a suitable reference frame, the geodesic deviation equation describes specific coordinate-independent gravitational effects experienced by a free observer in an arbitrary TMG spacetime. This is very useful for the physical interpretation.

The general form of the equation of geodesic deviation \eqref{Geodesic_Deviation_eq}, valid in any dimension and any geometry with the Levi-Civita connection, describes the effect of curvature on the second absolute derivative of the separation vector between two nearby geodesic particles. The dependence is encoded in the components of the Riemann tensor, which in TMG can be fully rewritten using the Cotton--York tensor $Y_{ab}$ and the energy-momentum tensor $T_{ab}$, as given by equation \eqref{Riemann_tensor_Cotton-York}. Reformulating the geodesic deviation in an arbitrary reference frame adapted to the observer's velocity $\boldu $ we derived a fully generic expressions \eqref{geodesic_deviation_frame_1}, \eqref{geodesic_deviation_frame_2}. It show that all the gravitational contributions can be characterized by the Cotton scalars $\Psi _A$, introduced in \cite{PP:2023,PP:2024}, and the cosmological constant.

Using this canonical decomposition of the geodesic deviation, in Section~4  we analysed the influence of individual gravitational components. Specifically we identified the isotropic effect \eqref{Influence_of_Lambda} of the cosmological constant~$\Lambda$, the transverse effects \eqref{Influence_of_Psi_0} and \eqref{Influence_of_Psi_4} of $\Psi _0$ and $\Psi _4$, the Newton-like influence \eqref{Interpret_eq_Psi_2} of $\Psi _2$, and the longitudinal effects \eqref{Influence_of_Psi_1} of \eqref{Influence_of_Psi_3} of $\Psi _1$ and $\Psi _3$.

In subsequent section~5, employing the Lorentz transformation we constructed an unique canonical interpretation frame based on the algebraic structure of a given spacetime \eqref{reciprocal_boost}, \eqref{null_rotation_fixed_k}, \eqref{rotation_parameter_L}. We then discussed particular contributions of spacetimes of a specific algebraic type to the equation of geodesic deviation, namely \eqref{geodesic_deviation_type_I_(1)}, \eqref{geodesic_deviation_type_I_(2)} for type~I, \eqref{geodesic_deviation_type_II_(1)},~\eqref{geodesic_deviation_type_II_(2)} for type~II, \eqref{geodesic_deviation_type_III_(1)}, \eqref{geodesic_deviation_type_III_(2)} for type~III, \eqref{geodesic_deviation_type_N_(1)}, \eqref{geodesic_deviation_type_N_(2)} for type~N, and \eqref{geodesic_deviation_type_D_(1)}, \eqref{geodesic_deviation_type_D_(2)} for type~D.

Moreover, in Section~6 we investigated the geodesic deviation effects caused by typical matter sources, namely pure radiation, perfect fluid, and a fully general Maxwell field.

Finally, we explicitly investigated the Brinkmann-like spacetimes \eqref{TN-wave_metric} in TMG, and we solved the equation of geodesic deviation for a special subclass of these spacetimes which admits a covariantly constant null vector field \eqref{pp-wave_metric}. For these pp-wave spacetimes, we showed that the curvature is determined by an arbitrary function of the retarded null coordinate~$u$, see \eqref{pp-wave_Psi_4_interp}. This is also the only Cotton scalar $\Psi _4 ^{\text{interp}}$ contributing to the equation of geodesic deviation in type~N spacetimes. When the wave-profile function in vacuum is a constant, the geodesic deviation depends only on its sign. For such a choice, we found an explicit solution for the separation of geodesics in terms of the Legendre functions \eqref{pp-wave_separation_vector_solutions}. The character of the solutions  is plotted in Figures~\ref{Plot_Z_f1>0} and~\ref{Plot_Z_f1<0}.

We hope that the general method developed in this paper will serve as a useful tool for future physical analyses of exact spacetimes in TMG. This may include investigation of a specific character of gravitational radiation with a massive graviton in the context of nonperturbative, exact theory of TMG.

\section*{Acknowledgments}

This work was supported by the Czech Science Foundation Grant No.~GA\v{C}R 23-05914S and by the Charles University Grant No.~GAUK 260325.

\newpage

\end{document}